\documentclass[journal=jacsat,manuscript=article]{achemso}
\usepackage{chemformula} 
\usepackage[T1]{fontenc} 

\usepackage{graphicx}
\usepackage{blindtext}
\usepackage{tabularx} 
\usepackage{array}
\usepackage{siunitx}
\usepackage{makecell}
\usepackage{tabularx}
\usepackage{mathtools}
\usepackage{lipsum}
\usepackage{amssymb}
\usepackage{textgreek}
\usepackage{amsmath}
\usepackage{multirow}
\usepackage{array}
\graphicspath{ {./pics/} }
\usepackage{etoolbox}
\author{Shishir Timilsena}
\affiliation{Department of Physics and Astronomy, University of Texas-Rio Grande Valley, Edinburg, TX 78539, USA}
\altaffiliation{These authors contributed equally to this work.}
\author{Dinesh Thapa}
\email{thapad@thomasmore.edu}
\affiliation{Department of Mathematics and Physics, Thomas More University, Crestview Hills, KY 41017, USA}
\alsoaffiliation{Department of Chemistry and Biochemistry, North Dakota State University, Fargo, North Dakota 58108, United States}
\altaffiliation{These authors contributed equally to this work.}
\author{David James Faller}
\affiliation{Department of Mathematics and Physics, Thomas More University, Crestview Hills, KY 41017, USA}
\author{Prabesh Adhikari}
\affiliation{Department of Nanoscience, Joint School of Nanoscience and Nanoengineering, University of North Carolina at Greensboro, Greensboro, North Carolina 27401, United States}
\author{Nicholas Dimakis}
\affiliation{Department of Physics and Astronomy, University of Texas-Rio Grande Valley, Edinburg, TX 78539, USA}
\author{Svetlana Kilina}
\affiliation{Department of Chemistry and Biochemistry, North Dakota State University, Fargo, North Dakota 58108, United States}

\title[An \textsf{achemso} demo]
  {Spin selective non-van der Waal electride nature in manganese under ambient pressure}

\abbreviations{IR,NMR,UV}
\DeclareUnicodeCharacter{0301}{\'{e}} 
\DeclareUnicodeCharacter{00E9}{\'{e}} 
\DeclareUnicodeCharacter{00F1}{\~{n}} 

\begin{document}


\begin{abstract}
 Electrides are an unusual class of ionic materials in which electrons localized in non-nuclear, interstitial regions act as anions within the crystal lattice. Here, we employ first-principles quantum mechanical calculations to investigate the structural, electronic, magnetic, and electride characteristics of elemental manganese (Mn) at an ambient pressure (0 GPa), focusing on its three crystalline phases: cubic ($\alpha$)-Mn ($I\bar{4}3m,no.217$), cubic ($\beta$)-Mn ($P4_132, no.213$), and hexagonal ($hex$)-Mn ($P6_3/mmc, no.194$). Our calculations reveal pronounced interstitial-electron character in all three phases, accompanied by spin-selective electron localization function (ELF), establishing elemental Mn as a non-van der Waals electride system. Bader charge analysis indicates substantial electron redistribution from the Mn host framework toward the interstitial anionic-electron (IAE) regions, with an effective charge transfer of approximately $-1.645e$, $-1.477e$, and $-1.083e$ per interstitial basin in $\alpha$-Mn, $\beta$-Mn, and $hex$-Mn, respectively. The electride character is further supported by the electronic density of states, where the IAE-associated states exhibit finite contributions near the Fermi level ($E_F$) and coexist with Mn-derived states, demonstrating their direct participation in the low-energy electronic structure. The combined electron localization function (ELF), effective charge transfer, and electron population due to IAE at $E_F$ therefore provide consistent evidence for interstitial anionic electrons in elemental Mn. To the best of our knowledge, this work provides the first systematic identification of spin-selective electride character in elemental Mn at ambient pressure, highlighting the possibility of exploiting its interstitial-electron states for unconventional electronic and magnetic functionalities.
\end{abstract}

\section{Introduction}
Electrides are the ionic solids that include anions in the form of cavity-trapped electrons, and remain sufficiently mobile to participate in conduction and chemical reactions.~\cite{dye2009electrides}. These loosely bound non-nuclear electrons give rise to special characteristics like high hyperpolarizabilities~\cite{xu2007structures}, low work functions~\cite{toda2008work}, and significantly high conductivity of various kinds~\cite{dye2003electrons}. 
Based on the environment in which anionic electrons are stabilized, electrides are broadly classified into layered van der Waal (vdW) electrides and framework/cage-type electrides (non-vdW electrides)~\cite{ liu2020electrides,zhou2024van}. In vdW electrides, electrons are confined in interlayer or interstitial positions within layered compounds in which electrons occupy two-dimensional (2D) interlayer spaces between positively charged atomic planes. When such electrides are exfoliated to a few atomic layers, the resulting structures are called electrenes, which show promise for applications in low-dimensional quantum electronics, including spintronics, valleytronics, and hydrogen energy storage~\cite{song2021van, zhou2024van}. In contrast, non-vdW electrides usually confine zero-dimensional (0D) pocket like or one-dimensional (1D) channel or tunnel like anionic electrons within three-dimensional (3D) structural voids, or interstitial sites of a rigid covalent or ionic framework, rather than in interlayer spaces held together by weak van der Waal forces ~\cite{rajput2025unveiling}.

Based on the dimensional distributions of interstitial anionic electrons (IAE), each group has distinct electrical and transport properties~\cite{tao2026accelerated}. Among these, the 2D electride Ca$_2$N has garnered a lot of interest because of its remarkable anisotropic magnetoresistance, low work function, and high electron mobility~\cite{lee2013dicalcium,druffel2016experimental}. Chang \textit{et al.}~\cite{liu2020electrides} presented a comprehensive classification of electrides into several categories, including organic, activated, high-pressure, native, magnetic, topological, and intermetallic systems. 
The IAEs impart distinctive characteristics to materials, including hyperpolarizabilities~\cite{srivastava2016m2x}, high magnetic susceptibilities~\cite{issa1984magnetic}, highly variable conductivities~\cite{hendrickson1998optical}, extremely low or ultra-low work functions~\cite{meng2021two,wang2022ultralow}, thermionic emission at cryogenic temperatures~\cite{phillips2000thermionic}, and very strong reducing capabilities~\cite{kim2015two}. These properties arise depending on the size, spatial arrangement, and nature of the trapped IAEs. Given their distinctive structure and characteristics, electrides are promising candidates for a wide range of potential applications, including superconductivity~\cite{hosono2021advances}, non-linear optical (NLO) response~\cite{srivastava2016m2x}, spintronics~\cite{sui2019prediction}, and electrocatalysis~\cite{weng2019theoretical}. A defining characteristic of electrides is their typically low  or ultra-low work function ($\Phi$) $\leq 3.9~ eV$ as illustrated in transition metal rich electride with partially filled $d$-orbitals\cite{wang2026electride, kang2022chemically, thapa2024signature}, and electride compounds with rare earth (RE) elements\cite{lee2020ferromagnetic}. Unlike low work function alkali metals, electrides provide a chemically and structurally tunable platform in which excess electrons are stabilized in crystallographically defined interstitial sites, enabling efficient electron donation, making electrides attractive for catalytic, electronic, and energy-related applications. Combining density functional theory (DFT) with experimental verification, Chung \textit{et al.} demonstrated that the Cu nanoparticles (NPs) when grown over the surface of $\rm Gd_2C$ electride, exhibits reduced work function of $\sim3.2 ~eV$, which is otherwise $4.5~eV$ in an ordinary Cu metal. The study confirmed that Cu nanoparticles supported on the low-work-function $\rm {[Gd_2C]^{2+}.2e^-}$ electride ($\sim2.8 ~eV$) can acquire excess electrons, which accumulate at the Cu surface and significantly enhance its oxidation resistance under ambient conditions. This approach suggests that electride-supported Cu nanoparticles could serve as stable, low-cost alternatives to noble-metal nanoparticles such as Au and Ag, with promising applications in catalysis and other electron-dependent technologies\cite{chung2022non}. Further, the $\mathrm{[Gd_2C]^{2+}\cdot2e^-}$ electride provides a unique platform for exploring quantum and topological phenomena associated with interstitial electrons. Kim \textit{et al.} identified a floating quantum electron liquid above the electride surface, exhibiting strong electron--electron correlations and signatures of non-Fermi-liquid behavior\cite{kim2022quantum}. Subsequent work by Lim \textit{et al.} revealed magnetic Weyl states and topological Fermi-arc surface states in the same two-dimensional electride. Remarkably, these topological states coexist beneath the floating quantum electron liquid, forming an unusual layered electronic structure at the surface\cite{lim2024topological}. Together, these studies reveal that electrides can host strongly correlated quantum electron liquids alongside nontrivial topological states, providing a promising platform for investigating emergent electronic phenomena. Recently, Adhikari and Thapa \textit{et al.} using computational approach, identified Ti$_2$N and Ti$_3$N$_2$ MXenes as a new class of electride materials due to quantum confinement of IAEs, with non-magnetic Ti$_2$N and antiferromagnetic (AFM) bulk Ti$_3$N$_2$ transitioning into a stable ferromagnetic (FM) electrene monolayer upon exfoliation. Park \textit{et al.}~\cite{park2024electride} demonstrated that electride formation induces superionic behavior in iron hydride (FeH) by stabilizing the Fe lattice and enhancing hydrogen diffusion under extreme pressures. Zhang \textit{et al.}~\cite{zhang2017electride} extended electrides to intermetallic compounds through Nb$_5$Ir$_3$, where interstitial electrons enabled tunable superconductivity with an enhanced transition temperature of \SI{10.5}{\K}. Sun \textit{et al.}~\cite{wang2026electride} highlighted the growing importance of electrides as heterogeneous catalysts for hydrogenation, carbon--carbon coupling, and electrocatalysis, while Miao \textit{et al.}~\cite{miao2014high} established a predictive framework for high-pressure electrides across the Periodic Table. Furthermore, Zhang \textit{et al.}~\cite{zhang2014two} identified Y$_2$C as a quasi-two-dimensional electride exhibiting semimetallicity, paramagnetism, and a low work function arising from interstitial electron--Y ($4d$) orbital hybridization.

Collectively, these studies establish electrides as a versatile class of quantum materials whose interstitial anionic electrons give rise to diverse phenomena, including magnetism, superconductivity, superionic transport, and catalytic activity. Their low work functions, tunable electronic structures, and exceptional electron-donating capabilities make electrides highly promising for future applications in spintronics, superconducting devices, heterogeneous catalysis, energy conversion and storage, and next-generation quantum technologies. Manganese (Mn), a ($3d$) transition metal with atomic number 25~\cite{serrano2026novel}, crystallizes in the complex body-centered cubic ($\alpha$)-phase below \SI{1000}{\K}, comprising 58 atoms per unit cell. This phase exhibits antiferromagnetic ordering and high electrical resistivity due to its intricate crystal structure~\cite{sliwko1994electronic,briere2002atomic,manago2022site,miyake2007electrical}. Above \SI{1000}{\K}, Mn transforms into the cubic ($\beta$)-phase with a 20-atom unit cell, accompanied by a transition from antiferromagnetic to paramagnetic behavior~\cite{sliwko1994electronic}. In this study, we have theoretically identified different stable phases of Mn (cubic-$\alpha$, cubic-$\beta$, and hexagonal $(hex)$) as native electride, where the zero-dimensional interstitial anionic electrons are an inherent part of the ground-state electronic structure of manganese. Since, Mn structures are stabilized in an antiferromagnetic (AFM) order, all the electronic, magnetic, and charge transfer properties in Mn lattice are explained in detail in this magnetic order.

\section{Computational Methods}

The first principle calculations were conducted within the framework of Density Functional Theory (DFT), as implemented in the Vienna Ab initio Simulation Package (VASP)~\cite{hafner1997vienna,hafner2008ab,sun2003performance}. The projector augmented wave (PAW) method and the Perdew–Burke–Ernzerhof (PBE) functional based on the generalized gradient approximation within the periodic boundary conditions (PBC), were employed to describe electron–ion interactions and exchange–correlation effects. A plane-wave kinetic energy cutoff of \SI{520}{\eV} was used throughout the calculations. The number of pseudopotential valence electrons ($N_{\mathrm{VE}}$) considered for Mn was 15, corresponding to the electronic configuration $3s^{2}3p^{6}4s^{2}3d^{5}$. It has been found that a larger value of $N_{VE}$ is suitable to get the actual value of effective charge transfer. The structural optimizations were performed with the Monkhorst-Pack of \(4 \times 4 \times 4\) k-point mesh for $\alpha-$ and $\beta-$ Mn, whereas \(18 \times 18 \times 18\) for $hex-$Mn until the forces on all atoms were less than \SI{0.01}{\eV\per\angstrom} and the total energy change between steps was below \(10^{-6}\) \si{\eV}. A Gaussian smearing of \SI{0.05}{\eV} was applied while calculating density of states (DOS). For the \(\alpha\)-Mn, $\beta$-Mn, and \(hex\)-Mn phases, Wigner–Seitz radii (\(R_{\mathrm{ws}}\)) of \SI{1.466}{\angstrom}, \SI{1.457}{\angstrom}, and \SI{1.156}{\angstrom} were assigned to Mn atoms, while empty spheres (pseudo atoms) were given \(R_{\mathrm{ws}}\) values of \SI{1.170}{\angstrom}, \SI{1.159}{\angstrom} and \SI{1.068}{\angstrom}, respectively with all \(R_{\mathrm{ws}}\) values derived from the atomic volumes obtained through Bader charge analysis~\cite{tang2009grid,sanville2007improved,henkelman2006fast,yu2011accurate}. The adjusted $R_{WS}$ value of Mn from Bader volume still lies well within or below the range of its covalent radii (1.39-1.61~\AA)\cite{phillips2000thermionic} confirming the atomic charge contribution do not overlap significantly with that of IAE. For the accurate description of the electronic and magnetic properties, the effective Hubbard parameter $U_{eff}=U-J=2.0 ~eV$ was used for the localized $d$ electron system of the Mn atom after several tests using Dudarev formalism \cite{dudarev2000correlation}. While $U_{eff}=0.0, 1.0$ and $1.5 ~eV$ substantially underestimate the total magnetic moment of the ferromagnetic (FM) phase of $hex-$Mn, $U_{eff}=2.0 ~eV$ gives the magnetic moment close to the hybrid functional HSE06 though it is known that the HSE06 sometimes overestimates the local magnetic moment on the transition metal due to large exchange splitting. Several other studies validate the choice of $U_{eff}=2.0 ~eV$ to treat the localized 3$d$ electrons in Mn to reproduce the experimental band gap reasonably well and to understand the fundamental magnetoelectric coupling.\cite{lane2023correlation, spurgeon2015polarization} The  Crystallographic visualizations, electron localization function (ELF), and magnetization densities were generated using the VESTA software~\cite{momma2011vesta}. The electron localization function (ELF) was decomposed into its spin-up ($\uparrow$) and spin-down ($\downarrow$) components to examine the spatial distribution and localization of anionic electrons in both spin channels. This decomposition was necessary because the default VESTA visualization displays only the spin-up component when the spin-resolved ELF data are derived from the total spin-unresolved ELF. Therefore, both spin channels were analyzed separately to provide a complete representation of the electron localization associated with the electride states.
\begin{figure*}[t]
    \centering
    \includegraphics[width=0.8\textwidth]{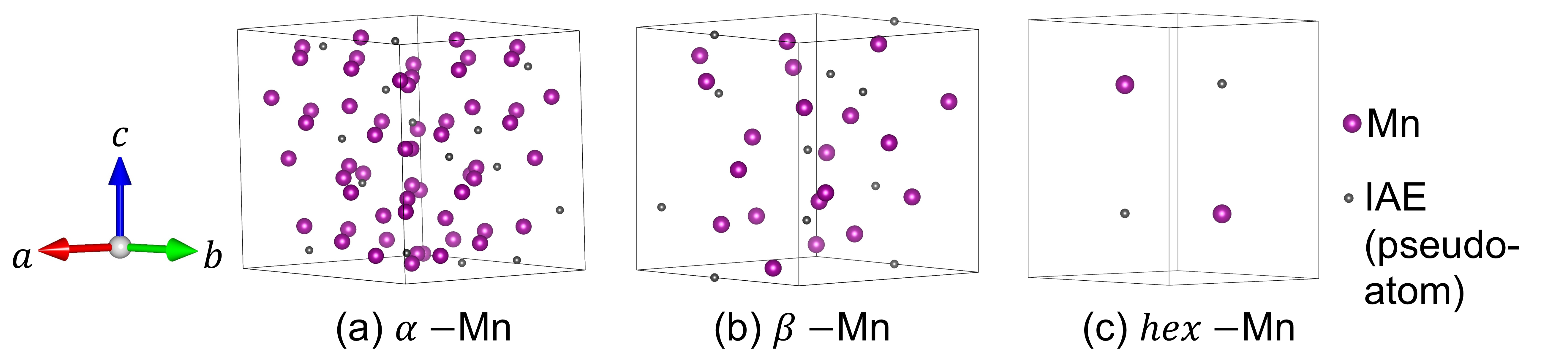}
    \caption{Optimized crystal structures of manganese (Mn) polymorphs: (a) $\alpha$-Mn, (b) $\beta$-Mn, and (c) $hex-$Mn.}
    \label{fig:Mn-structures}
\end{figure*}

\begin{figure}[htbp]
    \centering
    \includegraphics[width=0.4\textwidth]{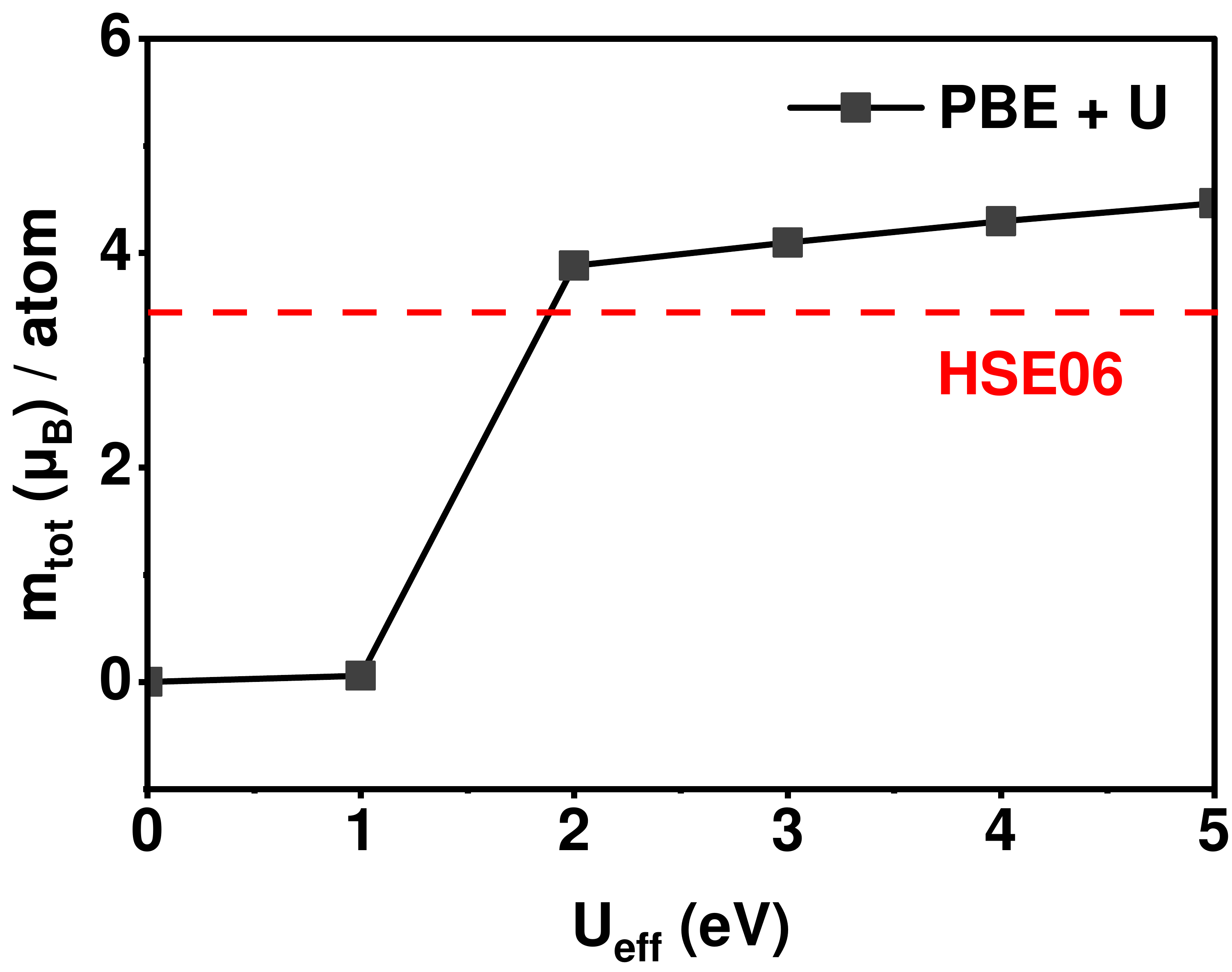}
    \caption{Total magnetic, $m_{tot}$ ($\mu_B$) per atom using standard PBE ($U_{eff}=0.0 ~eV$), $U_{eff}=1.0-5.0 ~eV$, and hybrid functional HSE06 in $hex-$Mn phase caluclated in ferromagnetic (FM) phase to benchmark effective Hubbard parameter.}
    \label{fig:Total-Magnetic-Moment}
\end{figure}

\section{Results and Discussion}
\subsection{Crystal structures}
The conventional crystal structures of cubic-$\alpha$-Mn, cubic-$\beta$-Mn, and hexagonal ($hex$)-Mn were obtained from the Materials Project~\cite{jain2013commentary} that belong to the space-group, $I\bar{4}3m (no.217)$, $P4_132 (no.213)$, and $P6_3/mmc (no.194$), respectively. The calculated lattice parameters ($a$, $b$, and $c$) and equilibrium volumes ($V_0$) for all three phases in ferromagnetic (FM) and antiferromagnetic (AFM) configurations are summarized in Table~\ref{tab:lattice_parameters}, while their optimized geometries, including the interstitial pseudo-atoms, are illustrated in Fig.~\ref{fig:Mn-structures}. The corresponding Wyckoff positions of the interstitial anionic electrons (IAEs) are provided in Supplementary Table~S1,S2 and S3. The calculated lattice parameters of the AFM phases are in good agreement with available experimental values for the corresponding AFM Mn phases. The optimized AFM structures exhibit small lattice distortions relative to their respective parent structures, with a more pronounced structural response in $\alpha$- and $\beta$-Mn than in $hex$-Mn. This phase-dependent relaxation is consistent with differences in spin--lattice coupling associated with AFM ordering, while the smaller response of $hex$-Mn suggests comparatively weaker coupling between its magnetic and structural degrees of freedom. These changes are therefore interpreted as magnetically induced structural relaxation within the respective parent crystal frameworks rather than as evidence of a change in crystallographic symmetry. Further, $\alpha$-Mn exhibits the shortest nearest-neighbor Mn--Mn bond length among the three phases, indicating stronger Mn--Mn bonding and providing a structural basis for its comparatively stronger lattice cohesion.
\begin{table}[htbp]
\caption{Calculated lattice parameters (in \AA) and equilibrium volume
(\AA$^3$) for different Mn phases. Here, Mn--Mn denotes the
shortest nearest-neighbor bond length. Experimental lattice parameters
for the cubic AFM phases are given in square brackets.}
\label{tab:lattice_parameters}
\centering
\begin{tabular}{lcccccc}
\hline
Structure & State & $a$ & $b$ & $c$ & $V_0$ & Mn--Mn \\
\hline
$\alpha$-Mn & AFM & 9.561 & 9.559 & 9.568 & 873.759 & 2.41 \\
            &     & [8.91]$^{\xi}$ & & & & \\
            & FM  & 9.875 & 9.875 & 9.875 & 962.906 & 2.58 \\
\hline
$\beta$-Mn  & AFM & 6.670 & 6.822 & 6.849 & 311.584 & 2.55 \\
            &     & [6.30]$^{\xi}$, [6.29]$^{\tau}$ & & & & \\
            & FM  & 6.926 & 6.926 & 6.926 & 332.245 & 2.61 \\
\hline
$hex$-Mn    & AFM & 2.857 & 2.857 & 4.399 & 31.112 & 2.75 \\
            & FM  & 2.834 & 2.834 & 4.575 & 31.795 & 2.81 \\
\hline
\end{tabular}

\vspace{2mm}
\begin{flushleft}
\footnotesize
$^{\xi}$Ref.~\cite{kasper1956antiferromagnetic};
$^{\tau}$Ref.~\cite{hafner2003understanding}.
\end{flushleft}
\end{table}

\subsection{Electronic properties}

To understand the electronic properties of the Mn electrides examined in this study, we employed the electron localization function (ELF), effective charge transfer, 
 electronic band structures, and density of states (DOS).

\subsubsection{Electron Localization Function}

The electron localization function (ELF), introduced by Becke and Edgecombe~\cite{becke1990simple}, provides a dimensionless real-space measure of the localization of same-spin electrons and has become a widely used descriptor for identifying chemically distinct regions such as covalent bonds, lone pairs, atomic cores, and interstitial electron localization in crystalline solids. The ELF at a spatial position $\mathbf{r}$ is defined as
\begin{equation}
\mathrm{ELF}_{\sigma}(\mathbf{r}) =
\frac{1}{1+\left[D_{\sigma}(\mathbf{r})/
D_{\sigma}^{0}(\mathbf{r})\right]^2},
\end{equation}
where $D_{\sigma}(\mathbf{r})$ represents the excess local kinetic-energy density associated with the Pauli exclusion principle for electrons of spin $\sigma$, and $D_{\sigma}^{0}(\mathbf{r})$ is the corresponding reference quantity for a homogeneous electron gas with the same local spin density. The quantity $D_{\sigma}(\mathbf{r})$ is expressed as
\begin{equation}
D_{\sigma}(\mathbf{r}) =
\tau_{\sigma}(\mathbf{r})
-\frac{1}{4}
\frac{|\nabla\rho_{\sigma}(\mathbf{r})|^2}
{\rho_{\sigma}(\mathbf{r})},
\end{equation}
where $\tau_{\sigma}(\mathbf{r})$ and $\rho_{\sigma}(\mathbf{r})$ are the positive-definite kinetic-energy density and electron density for spin $\sigma$, respectively. The first term is obtained from the occupied Kohn--Sham orbitals according to, $\tau_{\sigma}(\mathbf{r}) =
\sum_i |\nabla\psi_{i\sigma}(\mathbf{r})|^2$, while the second term corresponds to the von Weizs\"acker contribution. The homogeneous-electron-gas reference is given by, $D_{\sigma}^{0}(\mathbf{r}) =
\frac{3}{5}(6\pi^2)^{2/3}
\rho_{\sigma}^{5/3}(\mathbf{r})$. Consequently, $\mathrm{ELF}=0.5$ represents the localization of a homogeneous electron gas and therefore serves as a reference for a relatively delocalized metallic-electron distribution, whereas values approaching unity indicate enhanced localization relative to this reference. Conversely, low ELF values correspond to regions of weak localization and/or low electron density. Importantly, a high ELF value by itself does not establish an anionic state; the spatial position of the ELF maximum relative to the atomic sites and bonding regions must also be considered.

\begin{figure*}[htbp]
    \centering
    \includegraphics[width=1.0\textwidth]{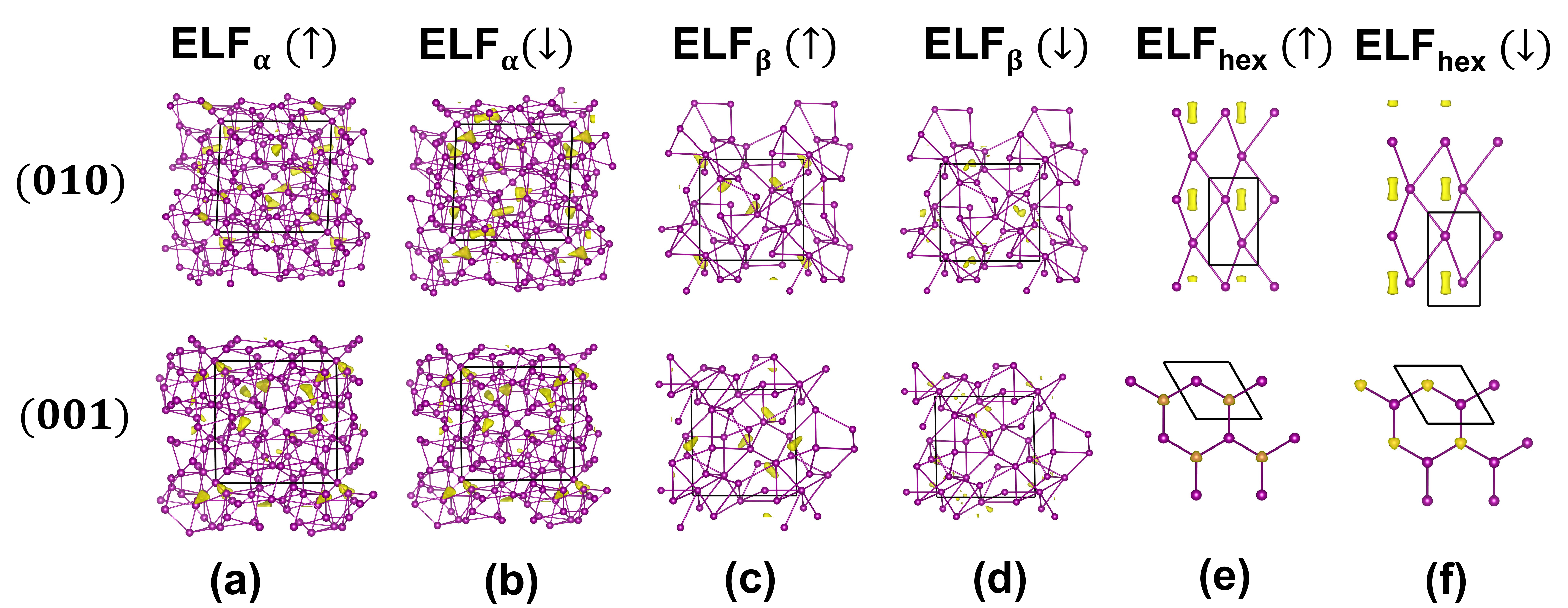}
    \caption{(a)--(f) 3D iso-surface plots represented by yellow balloon like structure of the electron localization function (ELF) with an iso-surface value of ELF = 0.68, illustrating the localization of interstitial anionic electrons (IAEs) for spin-up ($\uparrow$) and spin-down ($\downarrow$) channels separately in three different phases of Mn along the specified crystallographic planes (010)-side view and (001)-top view.}
    \label{fig:elf_3d_iso}
\end{figure*}

\begin{figure*}[htbp]
    \centering
    \includegraphics[width=1.0\textwidth]{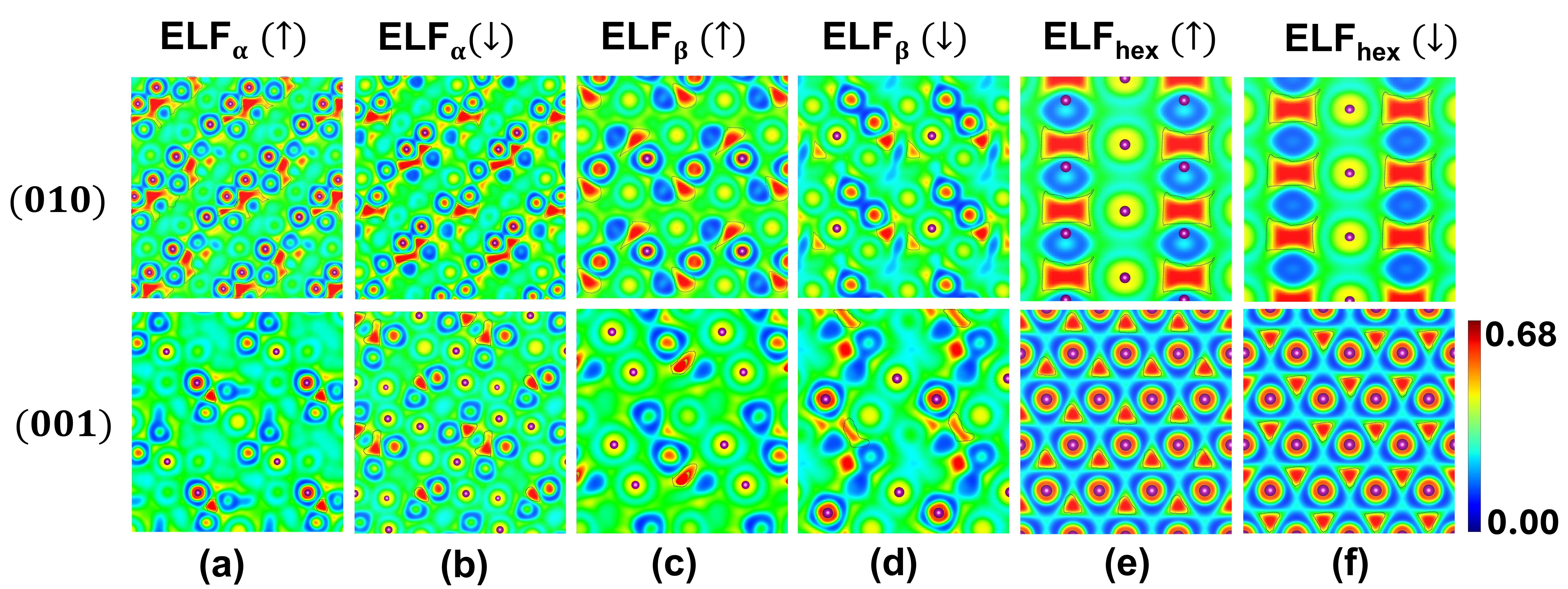}
    \caption{(a)--(f) 2D contour plots of the electron localization function (ELF) with ELF value ranges from 0.0 - 0.68, illustrating the localization of interstitial anionic electrons (IAEs) for spin-up ($\uparrow$) and spin-down ($\downarrow$) channels separately in three different phases of Mn along the specified crystallographic planes with slice plane passing through (010)-side view and (001)-top view. The in-plane localized ELF regions are highlighted by black dotted loops. 
    }
    \label{fig:elf_contours}
\end{figure*}

The calculated ELF distributions provide direct real-space evidence for pronounced electron localization at crystallographically defined interstitial sites in all three Mn phases. As shown in the three-dimensional iso-surface plots in Fig.~\ref{fig:elf_3d_iso}, pronounced ELF maxima are located away from the Mn nuclei and do not coincide with the conventional Mn--Mn bonding regions. Instead, these localized regions are centered within the voids of the Mn frameworks, identifying zero-dimensional (0D) electron-localization centers that are periodically distributed throughout the crystal. Such non-nuclear localization of electrons is a characteristic signature of electride materials, in which electrons occupying interstitial regions act as anionic species~\cite{tsirelson2002determination}. The observation of these localized ELF maxima in all three crystal structures therefore provides strong real-space evidence for the presence of interstitial anionic electrons (IAEs) in elemental manganese. The corresponding two-dimensional ELF contour maps in Fig.~\ref{fig:elf_contours}, obtained from the selected (010)-side view, and (001)-top view crystallographic planes, further resolve the spatial distribution of these interstitial localization centers. The high-ELF regions form closed, spatially confined contours within the voids of the Mn framework rather than being centered on the Mn nuclei. This spatial separation from the atomic cores and conventional Mn--Mn bonding regions is particularly important for identifying the localized electrons as interstitial species rather than as conventional bonding or core electrons. The periodic appearance of these localized regions across the crystallographic planes further demonstrates that the electron localization is an intrinsic feature of the crystal structures rather than an isolated numerical feature. To examine the spin properties of the interstitial electron localization, the ELF was resolved separately into spin-up ($\uparrow$) and spin-down ($\downarrow$) components. The resulting $\mathrm{ELF}{\uparrow}$ and $\mathrm{ELF}{\downarrow}$ distributions reveal localized interstitial regions in both the spin channels of ELF. The total number of IAE positions determined based on this localized ELF regions are 16, 8, and 2 within the unit cells of $\alpha$, $\beta$, and $hex$-Mn, respectively, which are often written as interstitial pseudo-atoms (empty spheres), dividing equally into ELF spin up (ELF$\uparrow$) and ELF spin down (ELF$\downarrow$). The presence of corresponding localization centers in the two spin channels indicates that the interstitial electron density is not exclusively associated with one spin orientation. Instead, the localized interstitial electrons participate in the overall spin-resolved electronic structure of the Mn lattice. This observation is relevant for the antiferromagnetic ground states of these phases, where opposite spin contributions can coexist spatially producing a small or vanishing net magnetic moment. The selected iso-surface value of $\mathrm{ELF}=0.68$ provides a convenient maximum visualization threshold for resolving the interstitial localization centers. At this value, the three-dimensional ELF display distinct iso-surfaces centered at non-nuclear positions within the Mn frameworks. The corresponding two-dimensional contours exhibit the same localized features, confirming that these regions persist across different crystallographic projections. The consistency between the three-dimensional iso-surfaces and two-dimensional contour maps demonstrates that the observed interstitial localization is spatially well defined and not an artifact of a particular viewing direction.

The ELF distributions also reveal a coexistence of localized and relatively delocalized electronic regions within the Mn lattices. The localized high-ELF regions are concentrated at the interstitial sites, whereas the surrounding regions exhibit lower ELF values characteristic of more delocalized electronic states. Such coexistence is consistent with the electronic nature expected for electrides, where localized interstitial electrons can coexist with more extended electronic states of the host framework. Therefore, the real-space ELF analysis establishes that the $\alpha$-, $\beta$-, and $hex$-Mn phases contain periodically arranged, non-nuclear electron-localization centers. The presence of these localized interstitial electron populations in all three structures at 0~K and 0~GPa provides compelling real-space evidence for electride character in the three Mn phases under study.

\subsubsection{Charge transfer analysis}

The charge-transfer behavior in the Mn phases was investigated using two complementary methods: (i) atomic charge analysis based on spherical integration within the PAW spheres in VASP, and (ii) Bader charge analysis~\cite{sanville2007improved, henkelman2006fast, tang2009grid}. The former provides a direct measure of the charge associated with the Mn atoms and the identified interstitial localization sites using the adjusted Wigner--Seitz radius, $R_{WS}$, whereas the latter independently partitions the total electron density into atom- and interstitial-centered Bader basins defined by zero-flux surfaces. Thus, the two approaches employ different spatial partitioning schemes but provide complementary measures of the same underlying charge redistribution within the Mn lattice. For the VASP-based atomic charge analysis, separate self-consistent (SC) calculations were performed using spherical integration regions with radii corresponding to the adjusted $R_{WS}$. The effective charge transfer is defined as
\begin{equation}
Q_{eff}^{(i)}=N_{VE}-Q_{tot}^{SC},
\end{equation}
where $N_{VE}$ is the number of pseudopotential valence electrons and $Q_{tot}^{(i)}$ is the integrated electron charge within the corresponding region calculated using method (i). For Mn, $Q_{tot}^{Mn}=Q_s+Q_p+Q_d$ whereas, because an IAE is a non-nuclear interstitial electron population rather than an atomic orbital state, its charge was obtained directly from the integrated electron density, $Q_{tot}^{IAE}=
\int_{\Omega_{IAE}}\rho(\mathbf{r})d^3r$, where $\Omega_{\mathrm{IAE}}$ is the chosen interstitial integration volume, defined as the spherical region centered at the IAE localization site. Here, $N_{VE}=15$ for Mn and $N_{VE}=0$ for the IAE. Consequently, positive and negative values of $Q_{eff}^{(i)}$ represent net charge depletion and accumulation, respectively. The VASP-based analysis yields average IAE charge accumulations of $-1.645e$, $-1.477e$, and $-1.083e$ per interstitial site, accompanied by Mn charge losses of $+0.953e$, $+1.045e$, and $+1.453e$ in $\alpha$-, $\beta$-, and $hex$-Mn, respectively. These results establish a quantitative charge redistribution from the Mn framework toward the interstitial localization sites.

As an independent validation of this charge-transfer picture, Bader analysis was performed using the pre-converged total charge density with reference to the augmented (core + valence) charge density. In contrast to the spherical integration used in the VASP analysis, the Bader method assigns electron density according to zero-flux surfaces of the charge-density gradient, thereby defining physically distinct atomic and interstitial basins~\cite{sanville2007improved}. The effective charge transfer was subsequently evaluated from the difference between the pseudopotential valence reference, $N_{VE}$, and the charge contained within the corresponding Bader basin, $Q_B$, calculated using method (ii) i.e.
\begin{equation}
    Q_{eff}^{(ii)}=N_{VE}-Q_B
\end{equation} 
Although the effective charge values obtained for Mn is significantly lower, particularly in $\alpha-$ and $\beta-$ Mn, both approaches give the consistent trend of electron depletion in Mn and electron accumulation at the interstitial sites. Similar to method (i), the Bader analysis also yields positive effective charge transfer for Mn and negative values for the IAEs, with Mn charge losses of $+0.404e$, $+0.515e$, and $+1.008e$ and IAE charge accumulations of $-1.464e$, $-1.289e$, and $-1.008e$ in the $\alpha$-, $\beta$-, and $hex$-Mn phases, respectively. 
The combined charge analyses therefore establish a consistent picture in which electronic charge is redistributed from the Mn atomic framework into the non-nuclear interstitial regions. Together with the pronounced interstitial localization observed in the ELF analysis, this charge accumulation provides quantitative support for the electride character of all three Mn phases. The Bader results should be interpreted as complementary to, rather than numerically identical with, the VASP spherical-integration results, since the two methods define the spatial boundaries of the electron populations differently. In particular, the ELF-defined interstitial sites provide an essential reference for identifying and characterizing the IAE basins, especially because interstitial electrons do not possess conventional atomic orbital assignments~\cite{thapa2024signature,song2021van}. The effective charge transfer in Mn and IAEs using both the methods are represented in Table:\ref{tab:Qeff}.


\begin{table}[htbp]
\caption{Effective atomic charges, $Q_{{eff}}^{(i)}$ and
$Q_{{eff}}^{(ii)}$, calculated using (i) self-consistent (SC)
VASP calculations and (ii) Bader analysis assuming a PAW sphere with
radius equal to the Wigner--Seitz radius ($R_{{WS}}$) for Mn
and interstitial anionic electrons (IAEs), as discussed in the
Computational Methodology section.}
\label{tab:Qeff}
\centering

\begin{tabular}{llrr|rr}
\hline
Phase & Species &
\multicolumn{2}{c|}{SC-VASP} &
\multicolumn{2}{c}{Bader Analysis} \\
\cline{3-6}
& &
$Q_{{tot}}^{{SC}}$ ($e$) &
$Q_{{eff}}^{(i)}$ ($e$) &
$Q_{B}$ ($e$) &
$Q_{{eff}}^{(ii)}$ ($e$) \\
\hline

$\alpha$-Mn & Mn
& 14.047 & +0.953 & 14.596 & +0.404 \\
& $IAE_{\mathrm{ave}}$
& 1.645 & $-1.645$ & 1.464 & $-1.464$ \\
& $IAE_{\mathrm{ELF}\uparrow}$
& 1.642 & $-1.642$ & 1.468 & $-1.468$ \\
& $IAE_{\mathrm{ELF}\downarrow}$
& 1.649 & $-1.649$ & 1.460 & $-1.460$ \\

\hline

$\beta$-Mn & Mn
& 13.955 & +1.045 & 14.485 & +0.515 \\
& $IAE_{\mathrm{ave}}$
& 1.477 & $-1.477$ & 1.288 & $-1.288$ \\
& $IAE_{\mathrm{ELF}\uparrow}$
& 1.406 & $-1.406$ & 1.381 & $-1.381$ \\
& $IAE_{\mathrm{ELF}\downarrow}$
& 1.549 & $-1.549$ & 1.196 & $-1.196$ \\

\hline

$hex$-Mn & Mn
& 13.547 & +1.453 & 13.992 & +1.008 \\
& $IAE_{\mathrm{ave}}$
& 1.083 & $-1.083$ & 1.008 & $-1.008$ \\
& $IAE_{\mathrm{ELF}\uparrow}$
& 1.083 & $-1.083$ & 1.003 & $-1.003$ \\
& $IAE_{\mathrm{ELF}\downarrow}$
& 1.083 & $-1.083$ & 1.013 & $-1.013$ \\

\hline
\end{tabular}
\end{table}

\subsection{Electronic Band structure and density of states}
\begin{figure}[htbp]
\centering

\includegraphics[width=0.35\textwidth]{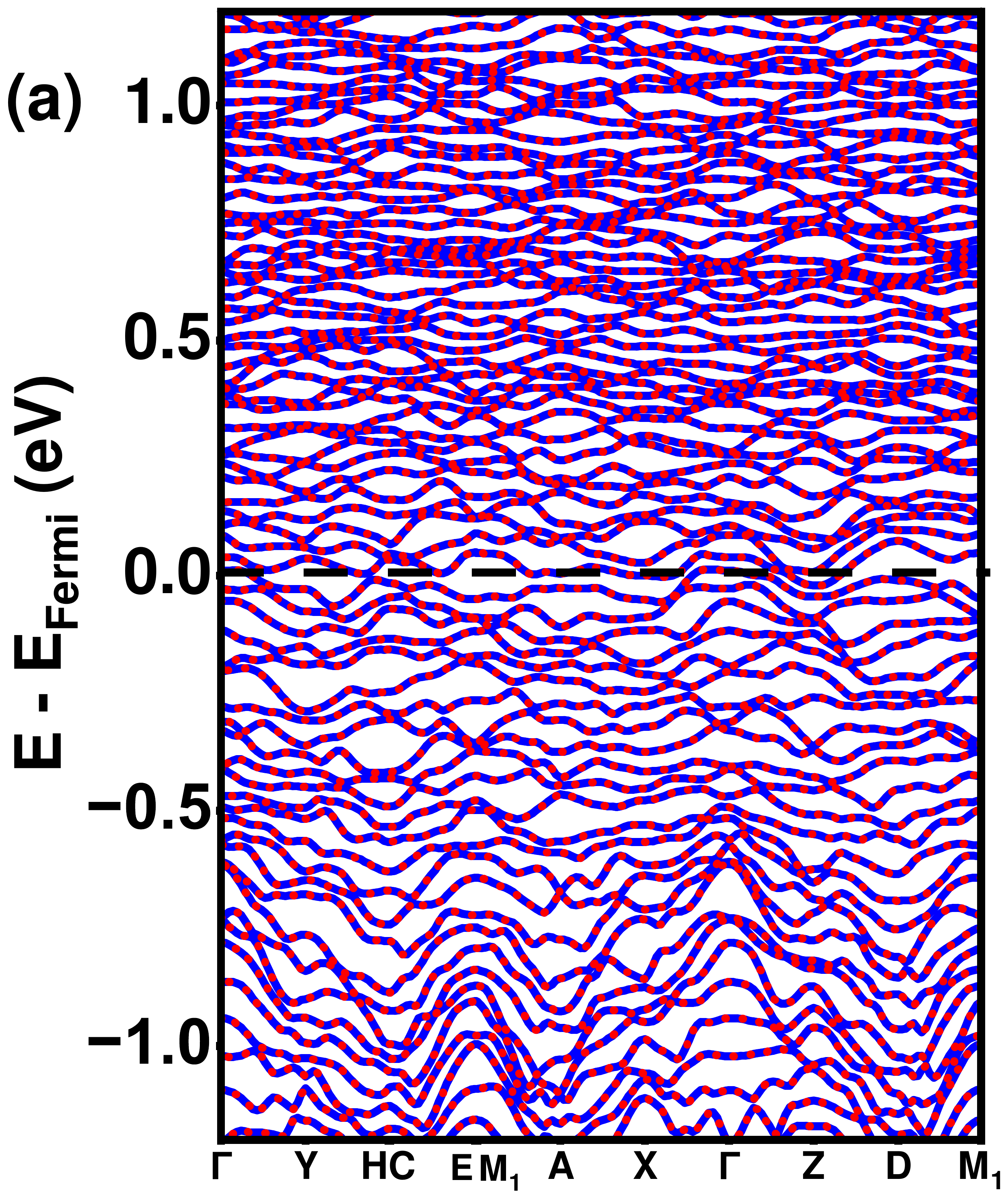}\\
\includegraphics[width=0.35\textwidth]{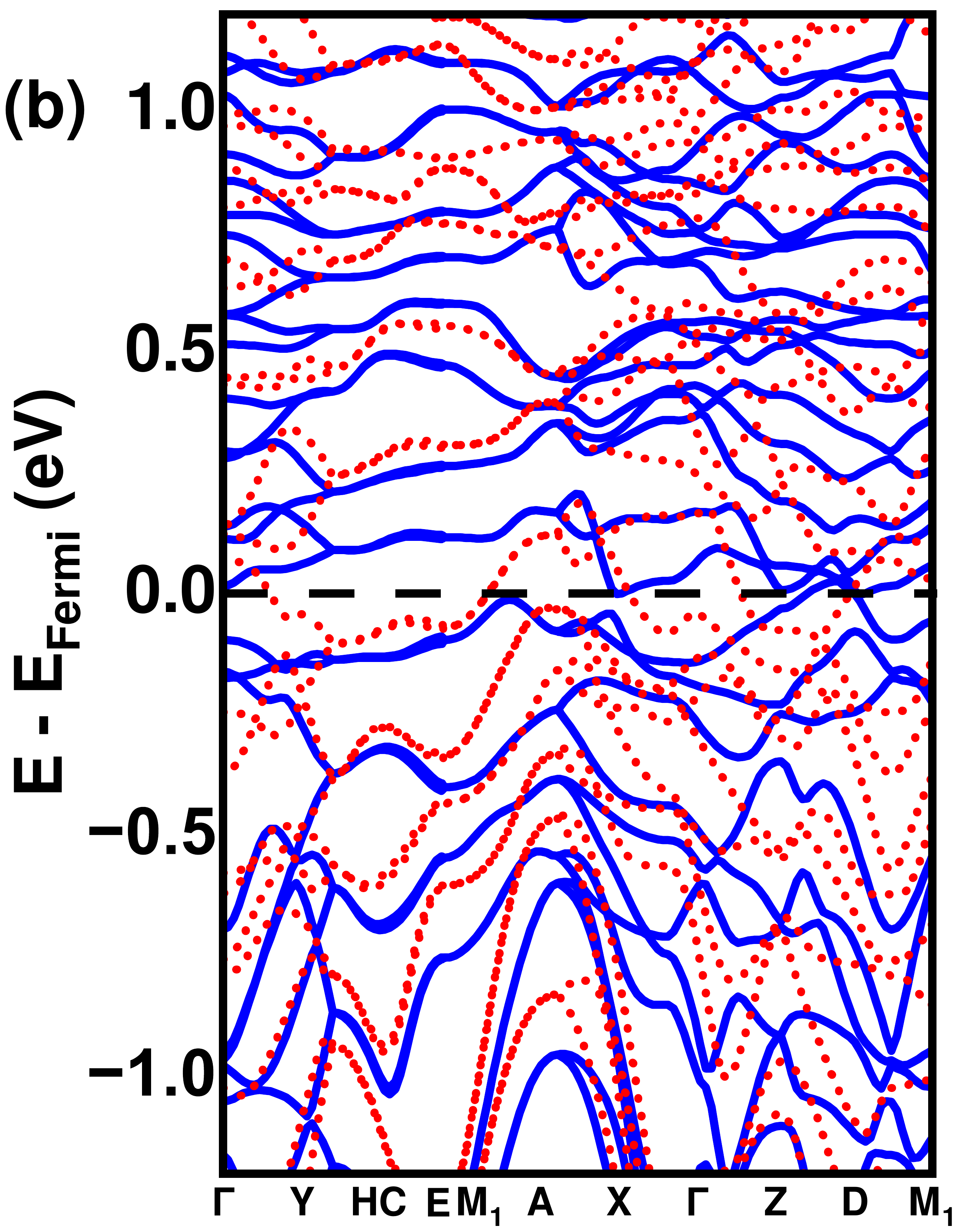}\\
\includegraphics[width=0.35\textwidth]{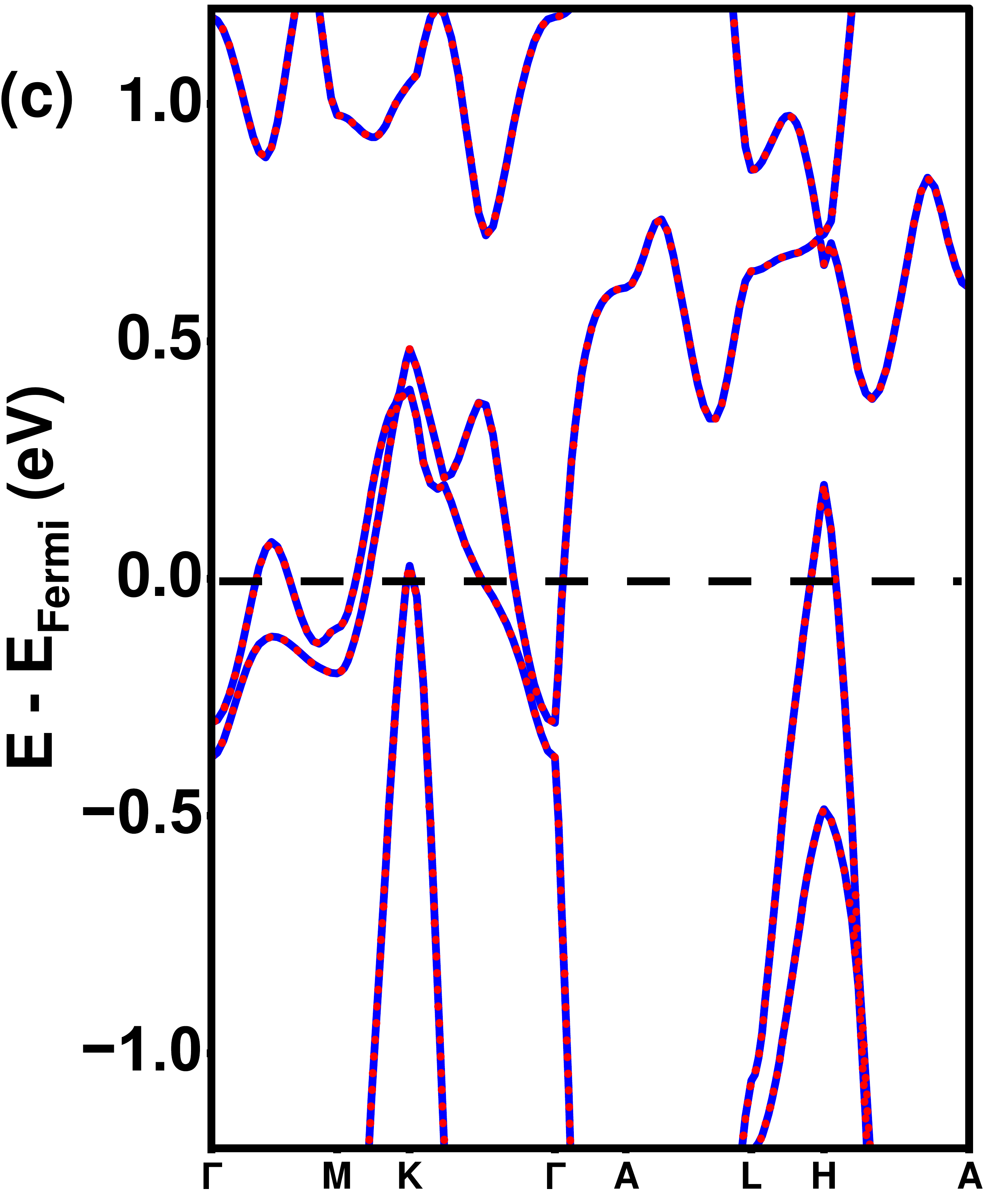}

\caption{Diagram representing total electronic band structure for
(a) $\alpha$-Mn, (b) $\beta$-Mn, and (c) $hex$-Mn. Here, the red dotted
lines and blue solid lines represent the spin-up and spin-down bands,
respectively. The dashed horizontal black lines in the energy axis
represent the Fermi level ($E_F$).}
\label{fig:band-structure}
\end{figure}

\begin{figure*}[htbp]
\centering
\includegraphics[width=1.0\textwidth]{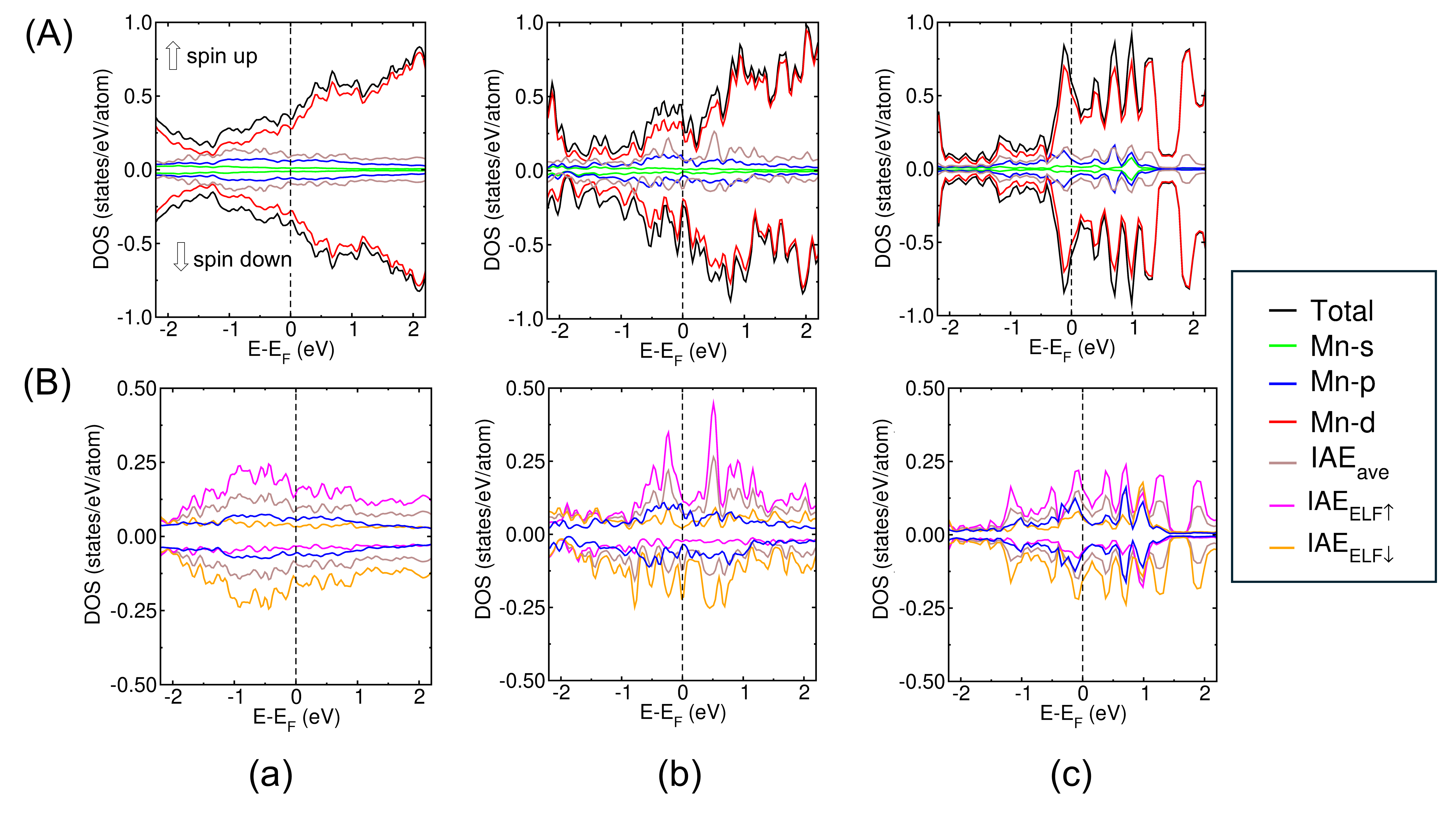}
\caption{Total and orbital projected density of states (DOS) for (a) \(\alpha\)-Mn, (b) \(\beta\)-Mn and (c) \(hex\)-Mn representing DOS comparison among (A) total, Mn-($s$, $p$,$d$) orbitals and average (ave) contributions from IAE per atom, (B) average contribution from IAE and pseudo-atoms located at spin up ($\uparrow$) and spin down ($\downarrow$) ELF region, denoted by $\rm IAE_{ELF\uparrow}$ and $\rm IAE_{ELF\downarrow}$. The dashed vertical black lines in the energy axis represent the Fermi level ($E_F$). }
\label{fig:dos}
\end{figure*}

The calculated electronic band structures and density of states (DOS) of the three Mn phases considered here consistently demonstrate their metallic character. As shown in Fig.~\ref{fig:band-structure}, the electronic bands cross the Fermi level ($E_F$), while the total DOS remains finite at $E_F$, i.e., $D(E_F)>0$, confirming the absence of a band gap in all three phases. The spin-up and spin-down states are nearly degenerate in the $\alpha$- and $hex$-Mn phases, whereas a slight lifting of spin degeneracy is observed in $\beta$-Mn, indicating a comparatively stronger spin polarization in this phase. To characterize the electronic contribution from the interstitial regions, the DOS associated with interstitial anionic electrons (IAEs) was obtained by introducing pseudo-atoms at the regions of strongly localized electron localization function (ELF), separately for the spin-up (ELF$\uparrow$) and spin-down (ELF$\downarrow$) regions. Notably, the resulting IAE states exhibit a substantial DOS contribution in the vicinity of $E_F$, exceeding the corresponding Mn-$s$ contribution and becoming comparable to the Mn-$p$ states. Moreover, the pseudo-atom projected DOS at the ELF$\uparrow$ regions shows a pronounced spin-up contribution relative to the spin-down contribution near $E_F$, whereas the opposite trend is observed for the ELF$\downarrow$ regions. This spin-resolved behavior provides evidence that the interstitial electrons are not merely localized charge accumulations but constitute electronically active states with appreciable spin character. Their significant spectral weight near $E_F$ therefore indicates that IAEs participate directly in the low-energy electronic states governing the metallic and spin-dependent electronic properties of these Mn phases. The electronic density of states comparing the DOS contributions from different orbitals ($s$, $p$, and $d$) of Mn, and the IAE located at $\rm ELF\uparrow$ and $\rm ELF\downarrow$ together with the average (ave) contributions of all the IAEs are shown in Fig.\ref{fig:dos}. 

To quantify the electronic contribution from the interstitial region, we further analyze the integrated density of states (IDOS) and compare the resulting electron populations with the intrinsic orbital contributions of Mn, as summarized in Table~\ref{tab:idos}. For a given atom or orbital, the IDOS up to the Fermi level, $E_F$ is defined as, $IDOS=N(E_F)=\int_{-\infty}^{E_F}D(E)\,dE,$ where $D(E)$ represents the corresponding projected DOS. The spin-resolved electron populations are then used to define the total electron contribution and spin polarization as, $\Delta N^{+}(E_F)
=
N^{\uparrow}(E_F)+N^{\downarrow}(E_F),$ and $\Delta N^{-}(E_F)
=
N^{\uparrow}(E_F)-N^{\downarrow}(E_F),$ respectively. Here, $\Delta N^{+}(E_F)$ represents the total number of electrons associated with a particular atom, orbital, or interstitial pseudo-atom up to $E_F$ and satisfies numerically, $\Delta N^{+}(E_F)\approx Q_{tot}^{SC}$, whereas $\Delta N^{-}(E_F)$ provides the spin imbalance and, when expressed in units of $\mu_B$, corresponds to the associated local magnetic moment. The magnetic moments obtained from the IDOS analysis, $\Delta N^{-}(E_F)$, show good agreement with the corresponding self-consistent magnetic moments, $m_{\mathrm{SC}}$ and satisfies numerically, $\Delta N^{-}(E_F)\approx m_{\mathrm{SC}}.$ The finite interstitial-derived DOS at the Fermi level, $D_{\mathrm{IAE}}(E_F)>0$, together with the substantial interstitial electron populations within the PAW spheres of IAE as, $\Delta N^{+}(E_F)\sim1.558e$, $1.422e$, and $1.030e$ for the $\alpha-$, $\beta-$, and $hex-$ Mn phases, respectively, provides direct evidence of the electron accumulation within the PAW spheres of the IAEs and their electronic activity. The effective charge transfers obtained from the PAW-sphere charge integration are in close agreement with those estimated independently from the orbital-resolved IDOS integrated up to the Fermi level, demonstrating consistent charge redistribution from the Mn framework toward the interstitial regions.The charge-transfer analysis provides further insight into the nature of Mn--Mn bonding in the three Mn phases. The effective charge depletion obtained from the IDOS analysis is $+0.965e$, $+1.093e$, and $+1.498e$ for $\alpha$-, $\beta$-, and $hex$-Mn, respectively, consistent with the trend obtained from the real-space PAW-sphere charge analysis. Similarly, the comparatively smaller charge depletion in $\alpha$-Mn, $+0.953e$(SC-VASP) or $+0.404e$(Bader program), indicates that a larger fraction of the Mn valence-electron density remains associated with the Mn framework. This behavior is consistent with the shorter Mn--Mn bond lengths in $\alpha$-Mn, which enhance the spatial overlap of neighboring Mn electronic states and thereby favor stronger Mn--Mn covalent bonding. The enhanced orbital overlap can promote greater localization of the bonding electron density within the Mn framework, reducing the extent of charge redistribution from the Mn sites toward the interstitial regions. In contrast, the longer nearest neighbor Mn--Mn bond lengths in $hex$-Mn, are accompanied by progressively larger effective charge depletion or more ionic character given by effective charge transfer of $+1.453e$ (SC-VASP) or $+1.008e$ (Bader Program), indicating greater redistribution of electron density away from the Mn-centered regions. Therefore, the correlation between shorter Mn--Mn bond lengths and smaller effective charge depletion provides consistent structural and electronic evidence for stronger Mn--Mn interactions in $\alpha$-Mn. 


\begin{table*}[htbp]
\caption{Spin-resolved integrated density of states (IDOS) and effective
charge transfer for Mn atoms and interstitial anionic electrons (IAEs)
in the $\alpha$-, $\beta$-, and $hex$-Mn phases. $N_{\uparrow}(E_F)$ and
$N_{\downarrow}(E_F)$ denote the integrated spin-up ($\uparrow$) and
spin-down ($\downarrow$) electron populations, respectively, at $E_F$.
$\Delta N^{+}(E_F)$ and $\Delta N^{-}(E_F)$ represent the total electron
population and spin-polarization contribution at $E_F$, respectively.
$Q_{{eff}}^{{IDOS}}$ is the effective charge transfer
obtained from the DOS analysis integrated up to $E_F$, defined as
$Q_{{eff}}^{{IDOS}}=N_{{VE}}-\Delta N^{+}(E_F)$ which numerically satisfies, $Q_{{eff}}^{IDOS}=Q_{{eff}}^{(i)}$, where $N_{{VE}}=15$ for Mn and $N_{{VE}}=0$ for IAE.
$m_{{SC}}$ is the local magnetic moment obtained from the
self-consistent (SC) calculation.}
\label{tab:idos}
\centering
\setlength{\tabcolsep}{7.2pt}
\renewcommand{\arraystretch}{1.0}

\begin{tabularx}{\textwidth}{
>{\centering\arraybackslash}l
>{\centering\arraybackslash}l
S[table-format=1.3]
S[table-format=1.3]
S[table-format=2.3]
S[table-format=+1.3]
S[table-format=+1.3]
S[table-format=+1.3]
}

\hline
\multicolumn{1}{c}{\rule{0pt}{2.8ex}Phase} & \multicolumn{1}{c}{\rule{0pt}{2.8ex}Orbital/IAE} & \multicolumn{1}{c}{\rule{0pt}{2.8ex}$N_{\uparrow}(E_F)$} & \multicolumn{1}{c}{\rule{0pt}{2.8ex}$N_{\downarrow}(E_F)$} & \multicolumn{1}{c}{\rule{0pt}{2.8ex}$\Delta N^{+}(E_F)$} & \multicolumn{1}{c}{\rule{0pt}{2.8ex}$Q_{{eff}}^{{IDOS}}$} & \multicolumn{1}{c}{\rule{0pt}{2.8ex}$\Delta N^{-}(E_F)$} & \multicolumn{1}{c}{\rule{0pt}{2.8ex}$m_{{SC}}$} \\ \multicolumn{1}{c}{} & \multicolumn{1}{c}{} & \multicolumn{1}{c}{(e)} & \multicolumn{1}{c}{(e)} & \multicolumn{1}{c}{(e)} & \multicolumn{1}{c}{$(e)$} & \multicolumn{1}{c}{$(\mu_B)$} & \multicolumn{1}{c}{$(\mu_B)$} \\
\hline

$\alpha$-Mn & $s$
& 1.220 & 1.220 & 2.440 & {} & 0.000 & 0.000 \\
& $p$
& 3.225 & 3.225 & 6.450 & {} & 0.000 & 0.000 \\
& $d$
& 2.572 & 2.572 & 5.145 & {} & 0.000 & 0.000 \\
& $s+p+d$
& 7.017 & 7.017 & 14.035 & +0.965 & 0.000 & 0.000 \\
& $IAE_{ave}$
& 0.779 & 0.779 & 1.558 & -1.558 & 0.000 & +0.001 \\
& $IAE_{ELF\uparrow}$
& 0.686 & 0.871 & 1.557 & -1.557 & -0.185 & -0.207 \\
& $IAE_{ELF\downarrow}$
& 0.874 & 0.686 & 1.560 & -1.560 & +0.188 & +0.208 \\

\hline

$\beta$-Mn & $s$
& 1.208 & 1.208 & 2.416 & {} & 0.000 & +0.002 \\
& $p$
& 3.211 & 3.199 & 6.410 & {} & +0.012 & +0.021 \\
& $d$
& 2.512 & 2.569 & 5.081 & {} & -0.056 & -0.025 \\
& $s+p+d$
& 6.931 & 6.976 & 13.907 & +1.093 & -0.044 & -0.002 \\
& $IAE_{ave}$
& 0.674 & 0.748 & 1.422 & -1.422 & -0.073 & -0.057 \\
& $IAE_{ELF\uparrow}$
& 0.531 & 0.816 & 1.347 & -1.347 & -0.285 & -0.280 \\
& $IAE_{ELF\downarrow}$
& 0.817 & 0.679 & 1.496 & -1.496 & +0.138 & +0.165 \\

\hline

$hex$-Mn & $s$
& 1.174 & 1.174 & 2.348 & {} & 0.000 & 0.000 \\
& $p$
& 3.120 & 3.120 & 6.240 & {} & 0.000 & 0.000 \\
& $d$
& 2.457 & 2.457 & 4.914 & {} & 0.000 & 0.000 \\
& $s+p+d$
& 6.751 & 6.751 & 13.502 & +1.498 & 0.000 & 0.000 \\
& $IAE_{ave}$
& 0.515 & 0.515 & 1.030 & -1.030 & 0.000 & 0.000 \\
& $IAE_{ELF\uparrow}$
& 0.435 & 0.595 & 1.030 & -1.030 & -0.159 & -0.173 \\
& $IAE_{ELF\downarrow}$
& 0.595 & 0.435 & 1.030 & -1.030 & +0.159 & +0.173 \\

\hline
\end{tabularx}
\end{table*}

\subsection{Ground state magnetic order and magnetization density}

The ground-state magnetic ordering of the $\alpha$-, $\beta$-, and $hex$-Mn phases is determined by comparing the total energies of the ferromagnetic (FM) and antiferromagnetic (AFM) configurations. The relative energy per atoms is defined as, $\Delta E = E_{{FM}}-E_{{AFM}}$, where $E_{{FM}}$ and $E_{{AFM}}$ denote the total energies of the FM and AFM configurations, respectively. Thus, $\Delta E>0$ indicates that the AFM configuration is energetically more favorable than the FM configuration. The calculated values of $\Delta E$ for the $\alpha$-, $\beta$-, and $hex$-Mn phases at ambient pressure (0~GPa) and 0 K are summarized in Table~\ref{tab:energy_comparison}. In all the three phases, the AFM configuration is found to be the ground-state magnetic ordering. The predicted AFM ground states of the $\alpha$- and $\beta$-Mn phases are consistent with the experimental neutron-diffraction results reported by Kasper and Roberts~\cite{kasper1956antiferromagnetic}. In the AFM ground state, the total magnetic moment of the crystal is nearly zero because the local magnetic moments on the Mn atoms are arranged in opposite spin orientations and compensate each other. Nevertheless, each Mn atom in average retains a substantial local magnetic moment of approximately $\pm 3.82~\mu_B$, $\pm 3.94~\mu_B$, and $\pm 3.85~\mu_B$ in the $\alpha$-, $\beta$-, and $hex$-Mn phases, respectively. Here, the positive and negative values correspond to the magnitudes in spin-up ($\uparrow$) and spin-down ($\downarrow$) orientations, respectively. The coexistence of large, oppositely aligned local moments and a nearly vanishing net moment is characteristic of the AFM ground state of these Mn phases.

\begin{table}[htbp]
\caption{Energy difference per atom between antiferromagnetic (AFM)
and ferromagnetic (FM) states, $\Delta E$, for manganese phases at
0 GPa and the corresponding ground-state magnetic order.}
\label{tab:energy_comparison}

\centering
\begin{tabular}{lcc}
\hline
Phase & $\Delta E$ (eV/atom) & \multicolumn{1}{c}{Ground-state} \\
      &                      & \multicolumn{1}{c}{magnetic order} \\
\hline
$\alpha$-Mn & 0.222 & AFM \\
$\beta$-Mn  & 0.148 & AFM \\
$hex$-Mn    & 0.065 & AFM \\
\hline
\end{tabular}
\end{table}

\begin{figure}
\centering
\includegraphics[width=0.7\textwidth]{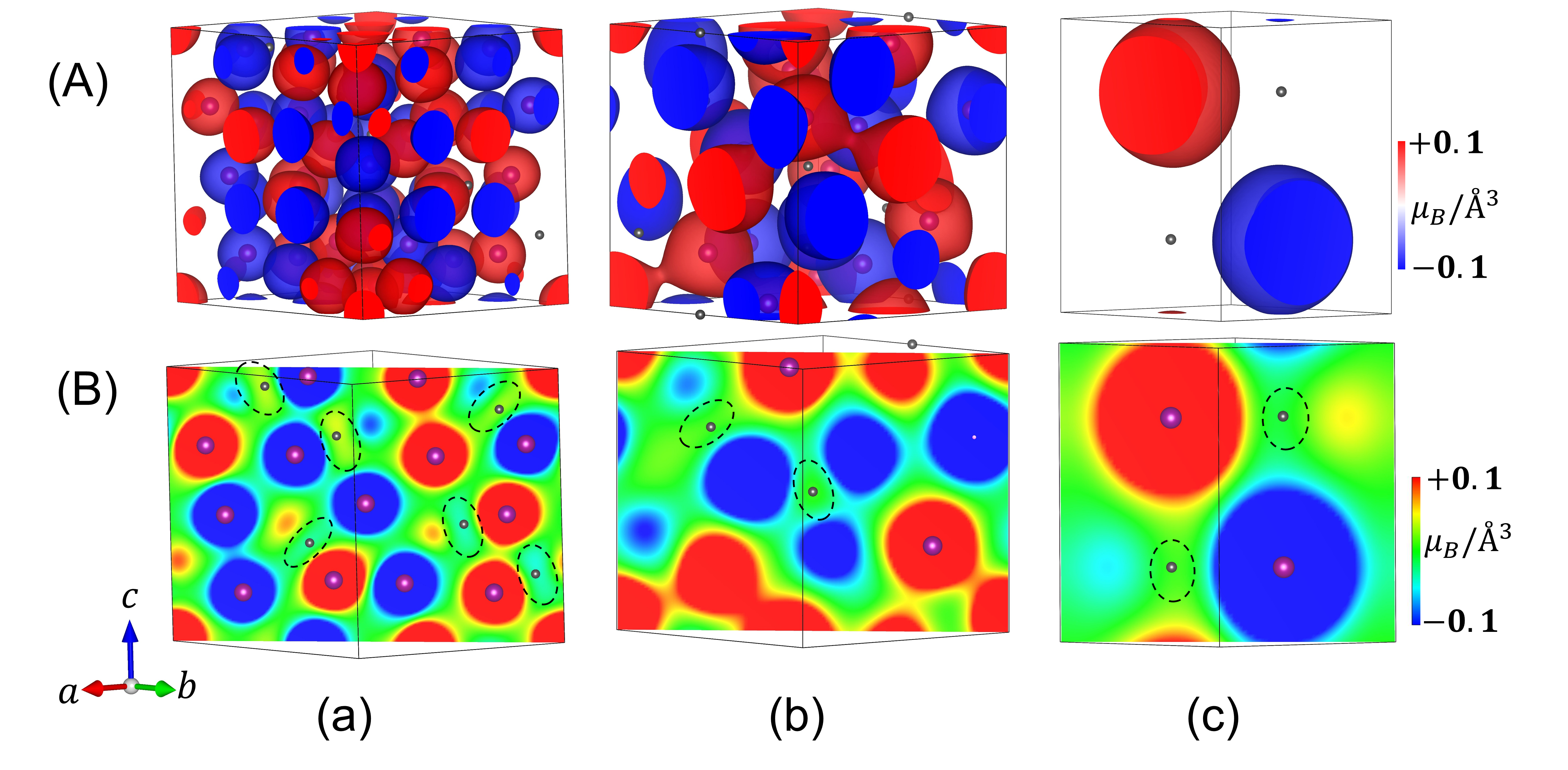}
\caption{(A) 3D iso-surface plots of magnetization density map (MDM) with iso-surface value $\pm0.1 ~\mu_B/$\AA$^3$, with red spin up ($\uparrow$) and blue spin down ($\downarrow$) density, and (B) 2D contour plots of MDM with slice plane passing through (110) plane for (a) $\alpha-$Mn, (b) $\beta-$Mn, and (c) $hex-$Mn. The weakly spin polarized IAE regions are encircled with dotted black loops. The color bar represents the range of MDM with red spin up and blue spin down density with green non-spin polarized region.}
\label{fig:magnetic_properties}
\end{figure}

The interstitial anionic electrons are identified from the pronounced electron localization at the interstitial sites and are represented by pseudo-atoms placed at the corresponding maxima of the spin-resolved electron localization function. Specifically, the IAE pseudo-atoms are positioned within the $\mathrm{ELF}{\uparrow}$ and $\mathrm{ELF}{\downarrow}$ regions, corresponding to the spatial locations where the interstitial electrons are predominantly localized in the spin-up and spin-down channels, respectively. As summarized in Table~\ref{tab:idos}, these IAE pseudo-atoms exhibit small but finite magnetic moments in all three Mn phases. The presence of IAE pseudo-atoms in both spin-up- and spin-down-localized ELF regions therefore indicates that the interstitial electron population is distributed among both spin channels. To distinguish the spatial localization of the interstitial electrons from their magnetic polarization, we further analyze the magnetization density, which is obtained from the difference between the spin-up ($\uparrow$) and spin-down ($\downarrow$) components of the total charge density ($\rho$)~\cite{lee2021mixed,thapa2025tunable}, $\rho_{{mag}}(\mathbf{r})
=
\rho_{\uparrow}(\mathbf{r})-\rho_{\downarrow}(\mathbf{r}).$ The resulting three-dimensional iso-surface plots of magnetization-density maps (MDMs) for the $\alpha$-, $\beta$-, and $hex$-Mn phases are presented in Fig.~\ref{fig:magnetic_properties}(A), where the red and blue iso-surfaces represent the spin-up ($\uparrow$) and spin-down ($\downarrow$) magnetization densities, respectively, over the selected iso-surface range of $\pm0.1e/$\AA$^3$. The MDMs show pronounced spin polarization localized around the Mn atoms, consistent with their large local magnetic moments. In contrast, the interstitial regions associated with the $\mathrm{ELF}^{\uparrow}$ and $\mathrm{ELF}^{\downarrow}$ maxima exhibit only weak magnetization density compared with the Mn sites. Thus, although the interstitial electrons are strongly localized and can be assigned to specific spin-resolved ELF regions, their net spin polarization is substantially smaller than that of the Mn $d$ electrons.

This behavior indicates that the interstitial electron population is predominantly spin compensated, with electrons occupying both spin channels, despite their spatially resolved localization in $\mathrm{ELF}{\uparrow}$ and $\mathrm{ELF}{\downarrow}$ regions. Consequently, the IAE states contribute significantly to the electronic structure and metallic character of the Mn phases but do not constitute the primary source of the large local magnetic moments. Instead, the dominant magnetic polarization originates from the Mn sites, while the interstitial electrons remain comparatively weakly spin polarized. Interestingly, the IAE at the ELF$\uparrow$ region exhibits spin down magnetic moment and that in ELF$\downarrow$ region exhibits spin up magnetic moment as shown in their numerical values in Table:\ref{tab:idos} and in 2D contour plots of MDM as shown in Fig.\ref{fig:magnetic_properties}(B). The coexistence of strongly localized, electronically active interstitial electrons with predominantly spin-compensated magnetization provides an important distinction between the electride character and the magnetic origin of these AFM Mn phases.

\section{Conclusion}

In summary, we have systematically investigated the structural, electronic, and magnetic properties of the three crystalline phases of elemental Mn, $\alpha$-, $\beta$-, and $hex$-Mn, using density functional theory (DFT) under periodic boundary conditions. Our results provide consistent evidence for electride character in all three phases. The calculated band structures and DOS demonstrate metallic behavior with finite density of states at $E$, while pronounced interstitial electron localization in the ELF identifies zero-dimensional anionic interstitial electrons (IAEs) spatially separated from the Mn atomic cores. Charge-transfer analysis using two independent approaches reveals substantial charge accumulation at the interstitial anionic electron (IAE) sites. Method (i), based on atomic charge analysis from self-consistent calculations, yields average IAE charge accumulations of $-1.645e$, $-1.477e$, and $-1.083e$ for $\alpha$-, $\beta$-, and hex-Mn, respectively. Method (ii), based on Bader charge analysis, gives corresponding charge accumulations of $-1.464e$, $-1.289e$, and $-1.008e$ per interstitial pseudo-atom for $\alpha$-, $\beta$-, and hex-Mn, respectively. The consistent negative charge obtained from both methods provides strong evidence for substantial electron accumulation at the interstitial sites in all three Mn phases.
Further, the substantial IAE contribution near $E_F$ and corresponding IDOS populations 
further demonstrate that these interstitial electrons are electronically active. Magnetic-energy comparisons show that the AFM configuration is energetically favored over the FM configuration in all three phases at 0~GPa, consistent with available experimental observations for $\alpha$- and $\beta$-Mn. Although the net AFM moment is nearly zero because of compensation between oppositely aligned Mn moments, substantial local moments of approximately $\pm3.82~\mu_B$, $\pm3.94~\mu_B$, and $\pm3.85~\mu_B$ per atom are retained in $\alpha$-, $\beta$-, and $hex$-Mn, respectively. Magnetization-density maps further show that spin polarization is predominantly localized around Mn, whereas the interstitial regions exhibit comparatively weaker spin polarization, with local moments $<0.2~\mu_B$ per interstitial atom. Importantly, these results distinguish the electronic origin of the electride state from the magnetic origin of the Mn phases. Although the IAE pseudo-atoms coincide with the corresponding spin-resolved ELF maxima, their magnetization density is substantially smaller than that of Mn, indicating predominantly spin-compensated interstitial electrons. Thus, strong interstitial electron localization does not necessarily imply substantial spin polarization, while the AFM ordering is predominantly governed by spin-polarized Mn $d$ states. The agreement between the spin-resolved IDOS-derived and self-consistent DFT magnetic moments further supports this interpretation.

Overall, $\alpha$-, $\beta$-, and $hex$-Mn simultaneously exhibit metallic electride behavior and antiferromagnetic ordering arising predominantly from distinct electronic populations. This coexistence of interstitial electron localization, metallicity, and robust local AFM moments establishes Mn as an unusual native magnetic electride under ambient pressure and provides a platform for exploring the interplay between interstitial-electron physics and magnetism, with potential implications for charge transfer and electrocatalytic phenomena.

\section{Conflict of interest}
The authors declare no conflict of interest. 

\section{Data and Code Availability}
All the data were produced using the Vienna ab-initio simulation package (VASP), a commercial software package. Most of the data produced are already mentioned in the maintext and Supplementary information in the form of tabular and diagrammatic representations.    

\section{Acknowledgements}
The authors gratefully acknowledge the computational resources provided by the Texas Advanced Computing Center (TACC) at The University of Texas at Austin. Authors also thank the Center for Computationally Assisted Science and Technology (CCAST) at North Dakota State University for providing computational resources used in this work, which were made possible in part by NSF EPSCoR RII Track-1: ND-ACES grant 1946202.

\subsection{Author Contribution}
Dinesh Thapa conceptualized, designed and supervised the project, performed DFT calculations, developed methodology, curated data, analyzed the results, implemented software, and developed computer codes. Shishir Timilsena performed DFT calculations, curated data, implemented software and analyzed the results. Dinesh Thapa and Shishir Timilsena wrote the original draft of the manuscript. All authors contributed actively in the formal analysis, validation, writing-review and editing.

\begin{suppinfo}
The Supporting Information (SI) is available free of charge at ... The SI includes the optimized POSCAR files together with the position coordinates of the IAEs in the stable antiferromagnetic (AFM) configuration of $\alpha-$, $\beta-$, and $hex-$ Mn phases.    

\end{suppinfo}

\bibliography{achemso-demo}

\end{document}